\documentclass[aps,prl,preprint,superscriptaddress]{revtex4-1}

\usepackage{graphicx}
\usepackage{subfigure}
\usepackage{color}
\usepackage{cancel}
\usepackage{amsmath}
\usepackage[linktocpage,colorlinks=true,linkcolor=blue,citecolor=blue,breaklinks=true]{hyperref}
\usepackage[normalem]{ulem}

\begin{document}


\title{Sparse game changers restore collective motion in panicked human crowds}
\author{Ajinkya Kulkarni}
\affiliation{Department of Applied Mechanics, Indian Institute of Technology Madras, Chennai 600036, India}

\author{Sumesh P. Thampi}
\email{sumesh@iitm.ac.in}
\affiliation{Department of Chemical Engineering, Indian Institute of Technology Madras, Chennai 600036, India}

\author{Mahesh V. Panchagnula}
\email{mvp@iitm.ac.in}
\affiliation{Department of Applied Mechanics, Indian Institute of Technology Madras, Chennai 600036, India}



\date{\today} 

\begin{abstract}
{Using a dynamic variant of the Vicsek model, we show that  emergence of a crush from an orderly moving human crowd is a non-equilibrium first order phase transition. We also show that this transition can be reversed by modifying the dynamics of a few people, deemed as game changers. Surprisingly, the optimal placement of these game changers is found to be in regions of maximum local crowd speed. The presence of such game changers is effective owing to the discontinuous nature of the underlying phase transition. Thus our generic approach provides (i) strategies to delay crush formation and (ii) paths to recover from a crush, two aspects that are of paramount importance in maintaining safety of mass gatherings of people.}
\end{abstract}



\maketitle


The Kumbh Mela in India \cite{baranwal2015, barnett2016} and the Hajj in Arabia \cite{hughes2003, Kim2015, curtis2013virtual} are the two biggest periodic human gathering events on earth. Estimates have shown that $\sim 10^6$ to $10^7$ people gather into a confined space during these events. The dynamics of such large crowds and its possible spontaneous transition to a crush have perplexed researchers for over thirty years. Gatherings at a carnival, in a sports stadium or at a train station are no less susceptible to crowd disasters \cite{kok2016}. Such transitions are a matter of grave concern to both law enforcement and public health. We study these phase transitions from an orderly movement to a crush in large confined mobile crowds with the motive of proposing a control strategy to delay the onset of a spontaneous transition to a crush or even reverse this state back to ordered motion.

The social force based model introduced by Helbing and Moln\`ar \cite{helbing1995} and its variants have been shown to be capable of describing the emergent dynamics in human crowds \cite{vicsek2000nature,kwak2013,Helbing2006prl,ma2013,cirillo2013}. Most of these previous studies have relied on a combination of theory and agent-based simulations to study escape dynamics and evacuation efficiency of crowds through narrow openings \cite{helbing1995}. Such approaches are now being complemented by on-site computer vision studies \cite{kok2016,Kim2015}, cognitive science \cite{moussaid2011} and data analytics \cite{helbing2007pre,Silverberg2013}. For example, deployment of authority figures in escaping crowds in a metro station \cite{Song2017}, placement of obstacles near the escaping door \cite{Lin2017,Zuriguelpre2016}, mixing individualistic and herding behavior \cite{vicsek2000nature} are being proposed as means to enhance the evacuation efficiency. Of course, analysis of these specific situations and mechanisms to avoid crowd disasters \cite{kwak2013,vicsek2000nature,ma2013,Chraibi2010} have led us to an understanding of the underlying dynamics. However, it is much more promising if we can generalize the physics and use this knowledge to devise strategies to control crowd behavior.

\begin{figure}
\centering
\includegraphics[height=3.2in]{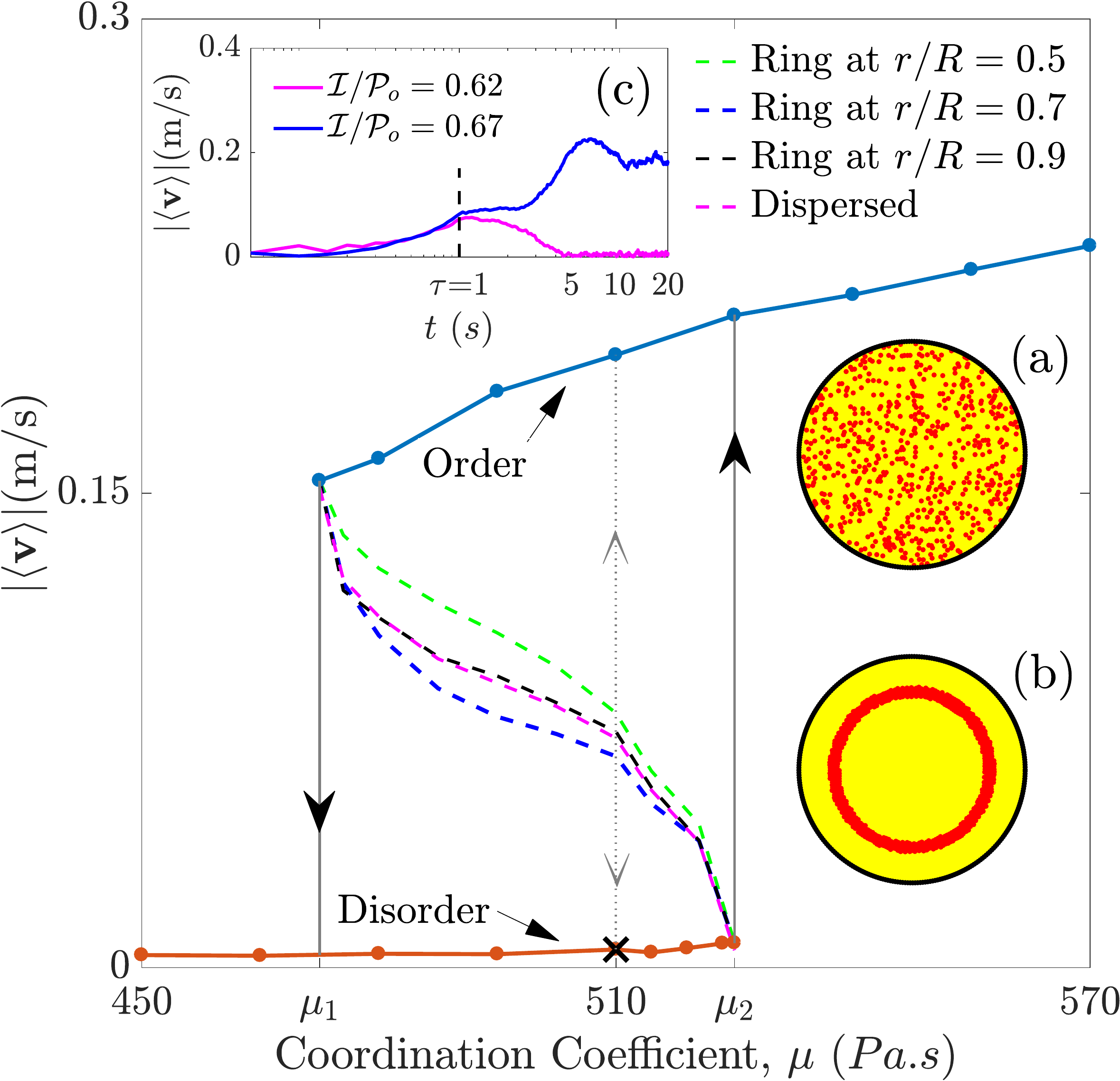}\label{fig:seperatrix}
\caption{Bifurcation diagram showing the states of dynamical order and disorder in a human crowd. As the coordination coefficient, $\mu$, between agents decreases, a uniformly moving crowd with a finite magnitude of the averaged velocity $\lvert \langle \mathbf{v} \rangle \rvert$ loses order and changes to a state of crush, with $\lvert\langle \mathbf{v} \rangle \rvert \approx 0$. Insets (a) and (b) show two different choices of location of game changers as red points among other people marked as yellow. Inset (c) shows a measure of success (finite $\lvert\langle \mathbf{v} \rangle \rvert$ as $t \rightarrow \infty$) and failure ($\lvert\langle \mathbf{v} \rangle \rvert \rightarrow 0$ as $t \rightarrow \infty$) for a given impulse $\mathcal{I}$ compared to the time-averaged momentum in the ordered state $\mathcal{P}_o$ for the value of $\mu$ marked with $\times$. The separatrices delineating domains of attraction towards the ordered and disordered states are shown by dashed lines.}

\label{fig:figure2}
\end{figure}


Indeed, in this \textit{Letter}, we show that such a generalized approach towards crowd control is possible.
Fig.~\ref{fig:figure2} summarizes the main findings of this work.
We model human crowd in the frame work of active matter \cite{marchetti2013rmp1,vicsek1995prl,vicsek2000nature} and obtain two distinct crowd states, namely an ordered state of collective motion and a disordered state of crush. The ordered state with a finite magnitude of the averaged velocity $\lvert\langle \mathbf{v} \rangle \rvert$ is realized when coordination amongst the agents is high. Here, $\langle \hspace{0.05cm} \cdot \hspace{0.05cm} \rangle$ denotes average over all agents and over time. A transition to a disordered state where $\lvert\langle \mathbf{v} \rangle \rvert \approx 0$ is observed as a result of breakdown in agent coordination. We will show that transition from collective ordered motion to a state of crush happens as a non-equilibrium phase transition. More importantly we find that this transition can be reversed, without requiring a change in the overall level of coordination, by imparting an impulse to a small fraction of randomly chosen agents deemed as \textit{game changers}. As we will show, game changers work best when they are placed in regions of maximum crowd speed (or lowest local panic). For example, the optimal location of game changers is on a ring located at 70\% of the radius in a circular domain. We now describe the model, the implications of these findings for the safety of a human crowd and the general rationale for optimal game changer placement.


 

\textit{Model description}:
We use a dynamical variant of the well-studied agent based Vicsek model \cite{vicsek2000nature, vicsek1995prl} to simulate crowd dynamics.
This is an off-lattice model, where each agent is modeled as having a mass $m$ and occupying an area of a soft disc of diameter $d$. $N$ such agents are confined to move in a circular boundary. Each agent responds to three isotropic `interaction forces'. Thus the momentum balance for an agent $i$ can be written as
\begin{align}
m \frac{d\mathbf{v}_{i}}{dt} = \mathbf{F}^{pp}_{i} + \mathbf{F}^{sp}_{i} + \mathbf{F}^{D}_{i} 
\label{eqn:forcebal}
\end{align}
where $\mathbf{F}^{pp}_{i}$, $\mathbf{F}^{sp}_{i}$ and $\mathbf{F}^{D}_{i}$ represent, respectively, the sum of all repulsive forces on the $i^{th}$ agent due to binary interactions with neighboring agents, a self propelling force generated by the $i^{th}$ agent and an alignment force on the $i^{th}$ agent due to agent-neighbor interactions. Agent inertia, which is not part of the classical Vicsek model \cite{vicsek1995prl} is included in Eq.~\eqref{eqn:forcebal}. The repulsive interaction force between the agent $i$ and a neighboring agent $j$ occurs due to space exclusion and is modeled as a linear soft spring: $\mathbf{F}^{pp}_{i}=\sum_j -k_n \boldsymbol{\delta}_{ij} \mathcal{H}\left( |\mathbf{r}_{i} - \mathbf{r}_{j}|-d \right)$.
Here, $\mathcal{H}(\cdot)$ is Heaviside function. If the position vector $\mathbf{r}_i$ points to the center of the $i^{th}$ agent, the extent of compression, $\boldsymbol{\delta}_{ij}$, is the vectorial distance along the separation vector between two agents $i$ and $j$,
$\boldsymbol{\delta}_{ij} = \left(|\mathbf{r}_{i} - \mathbf{r}_{j}|-d\right) \frac{\mathbf{r}_{i} - \mathbf{r}_{j}}{|\mathbf{r}_{i} - \mathbf{r}_{j}|}$. The coefficient $k_n$ determines the strength of the space exclusion force drawing from the analogous Herztian contact theory. The self propelling force $\mathbf{F}^{sp}_{i}$, produced by an agent has two components: one component aligned with the instantaneous velocity direction $\hat{\mathbf{v}}_{i}$ of strength $\beta$ and a second along a pre-determined motive direction $\hat{\mathbf{v}}^{m}_{i}$ of strength $\gamma$. Mathematically, $\mathbf{F}_{i}^{sp} =  m(\beta\hat{\mathbf{v}}_{i} + \gamma\hat{\mathbf{v}}^m_{i}) -\alpha |\mathbf{v}_{i}|\hat{\mathbf{v}}_{i}$ where $|\hat{\mathbf{v}}_{i}| = |\hat{\mathbf{v}}^m_{i}| = 1$. The parameter $\alpha$ is responsible for limiting the agent  to a terminal speed \cite{hinz2014motility}. The last force in the list is the collective crowd influence force experienced by each agent and is given by $\mathbf{F}^{D}_{i} =- \mu d (\mathbf{v}_{i} - {\mathbf{v}}_\mathbf{c})$, where $\mu$ is a co-ordination coefficient that controls the coupling strength between the $i^{th}$ agent and its neighborhood crowd. We calculate ${\mathbf{v}}_\mathbf{c}$ from a Gaussian weighted average of the velocities of all neighborhood agents in a radius $h$ \cite{pavan2015granular,Mahapatra2017}, chosen appropriately for dense crowds \cite{curtis2013virtual,Bechinger2016}.


We consider $N=6120$ active agents of diameter $d=0.5~m$ confined in a domain of circular shape of radius $R=22.5~m$ and mass $m=60~kg$, which results in a crowd density $\rho \approx 4~$persons$/m^2$ as observed in typical crowd conditions \cite{Parisi2009}. We select $k_n = 3 \times 10^{6}~N/m $ and $\beta = 1~m/s^{2}$. We have set $\alpha=\gamma=0$ without loss of generality of the conclusions. The conclusions also remain unaltered for systems with polydispersed agents \cite{sm1}. Typical agent speeds in the simulations match with those observed in literature \cite{Baglietto2011,helbing2007pre}. The influence radius for the neighborhood is set as $h=5d$ with the walls modeled as fixed agents. 

\begin{figure}
\centering
\subfigure[]{\includegraphics[height=0.93in]{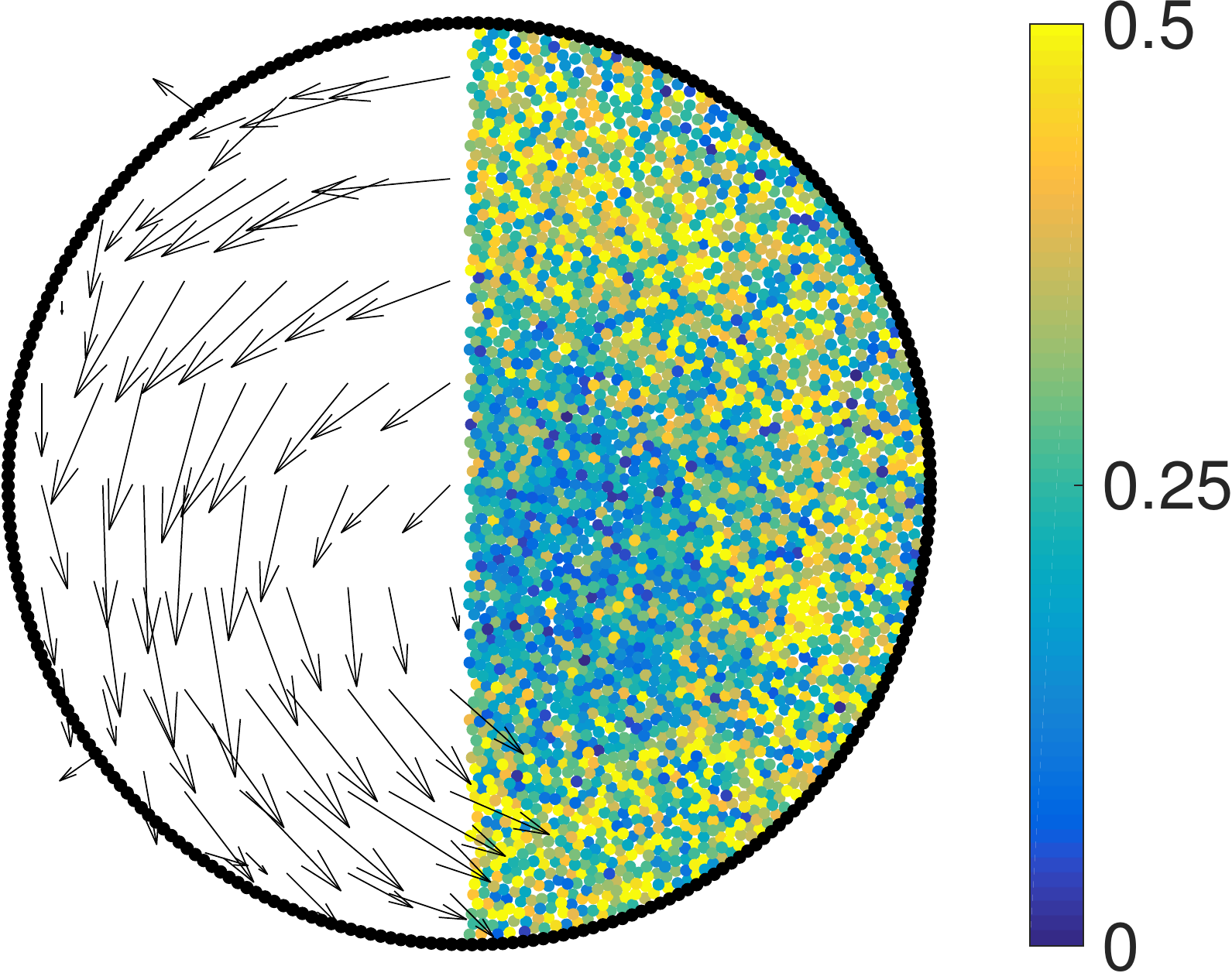}\label{fig:half3p6}}
\subfigure[]{\includegraphics[height=0.93in]{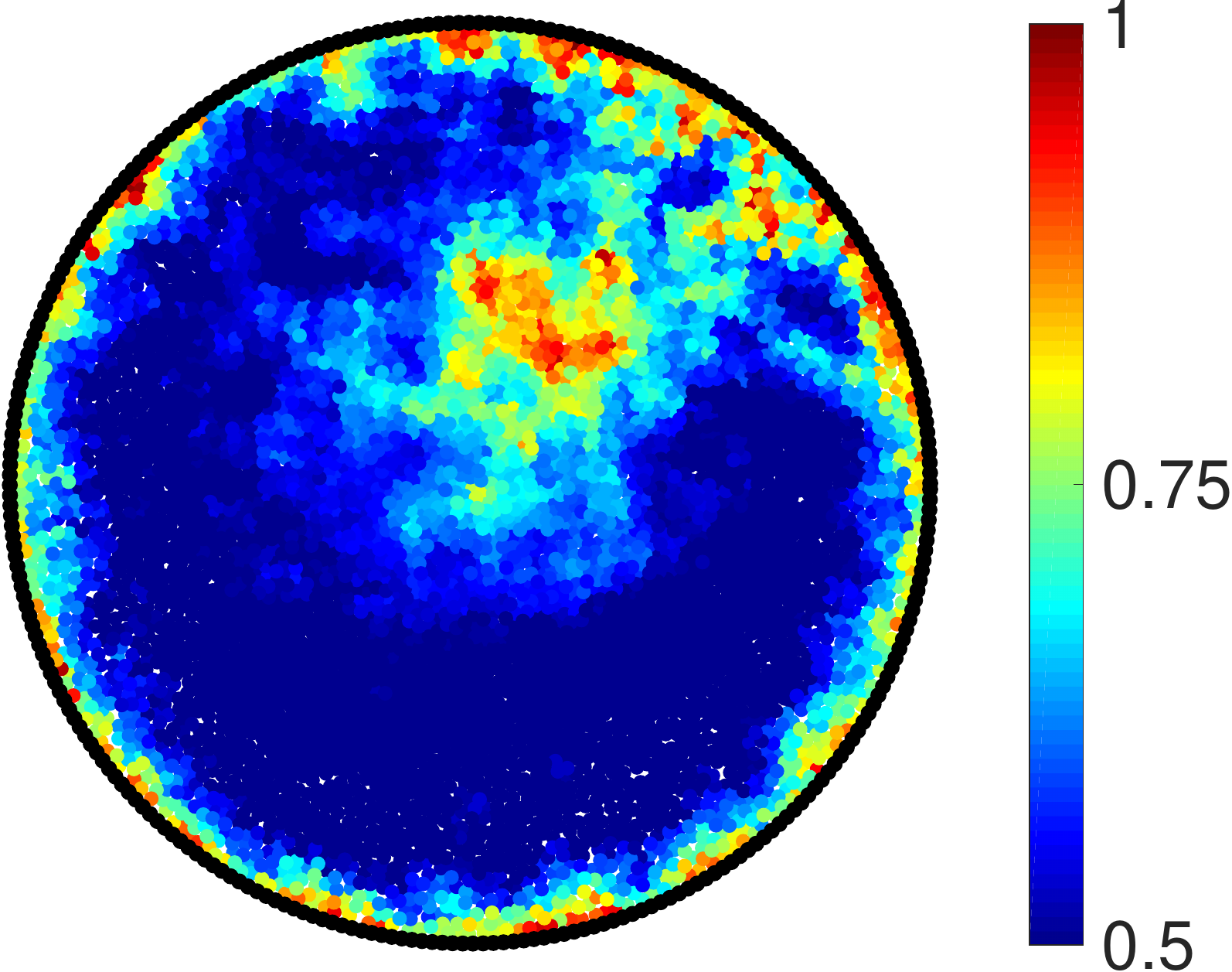}\label{fig:panic3p6}}
\subfigure[]{\includegraphics[height=0.95in]{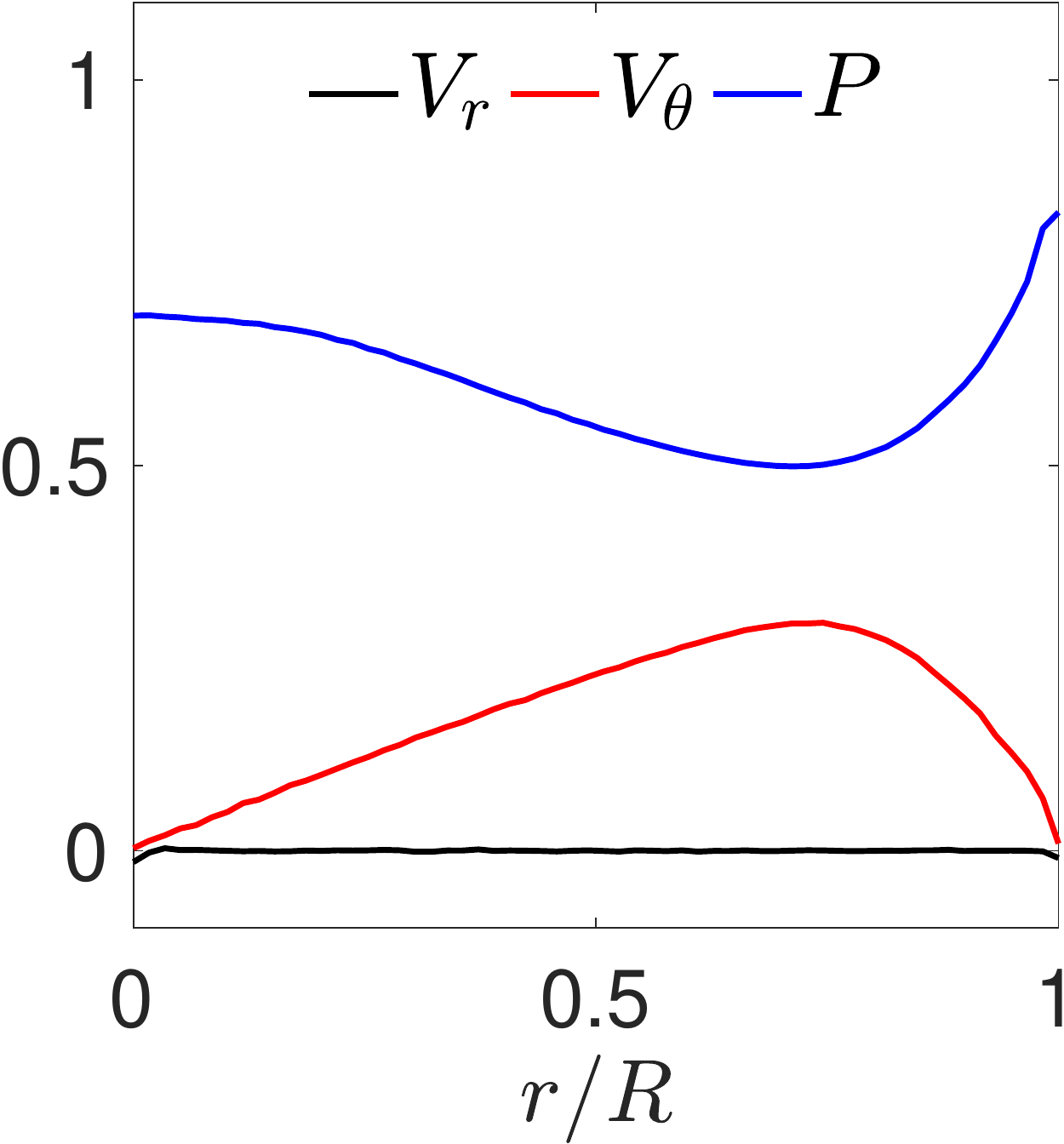}\label{fig:speed3p6}}\\
\centering
\subfigure[]{\includegraphics[height=0.93in]{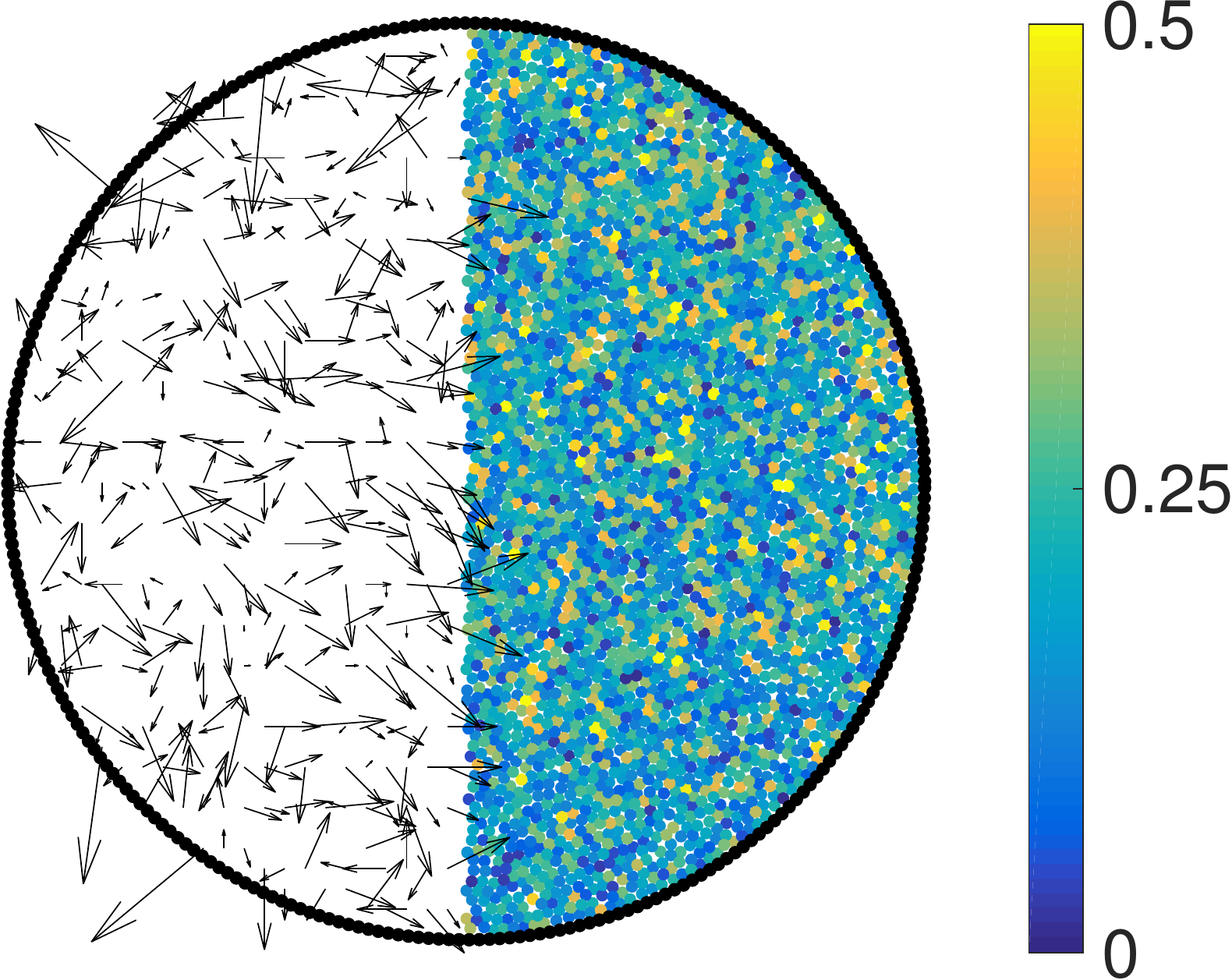}\label{fig:half3p4}}
\subfigure[]{\includegraphics[height=0.93in]{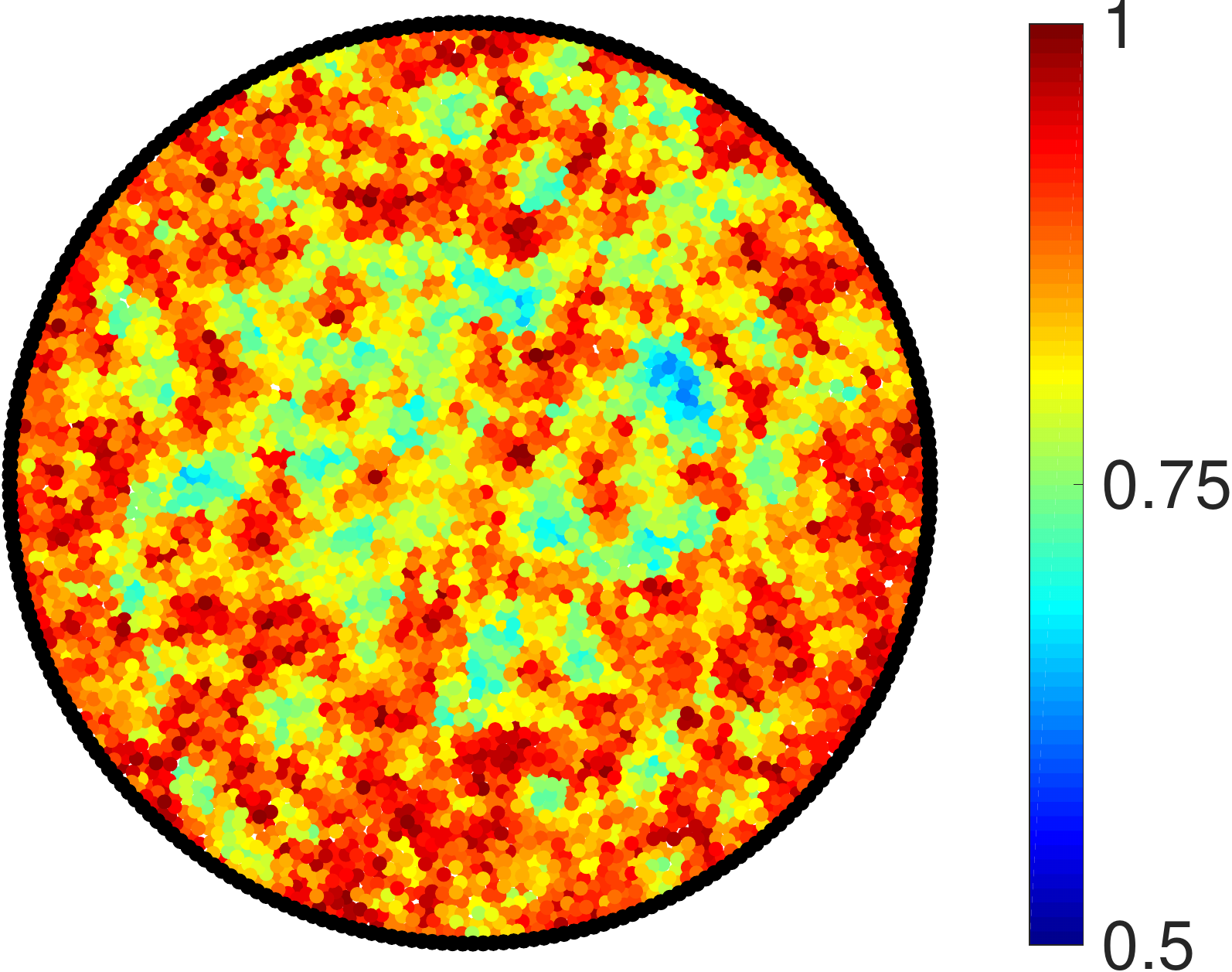}\label{fig:panic3p4}}
\subfigure[]{\includegraphics[height=0.95in]{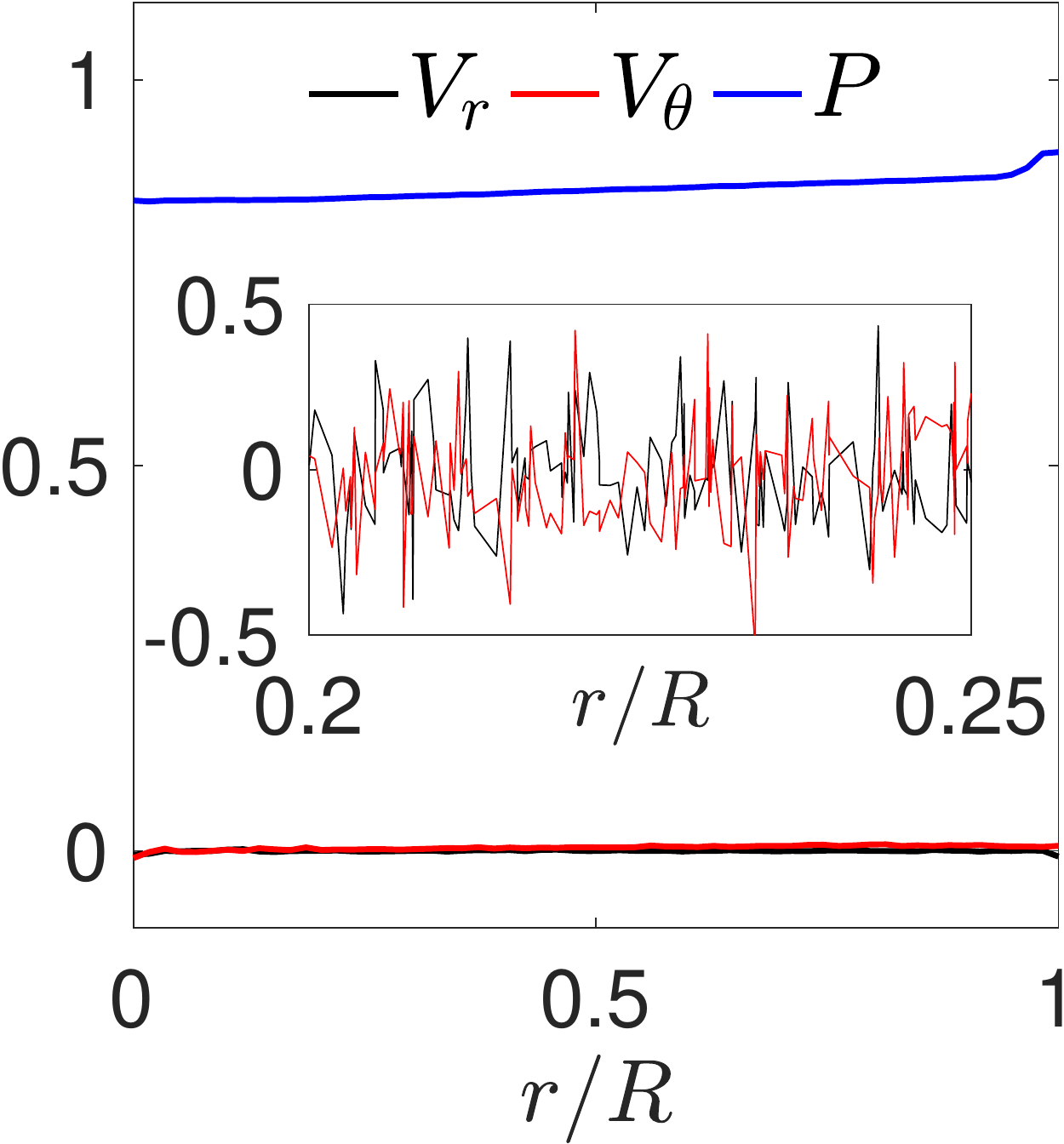}\label{fig:speed3p4}}
\caption{Instantaneous fields of (a) velocity of agents, left half shows direction and right half shows the magnitude (b) panic factor and the (c) radial distribution of azimuthally averaged radial and angular components of velocity and panic factor $P$ for a system exhibiting ordered motion at $\mu=540~(Pa\cdot{s})$.  As $\mu$ decreases to $510~(Pa\cdot{s})$, the system transits to a disordered state. Correspondingly (d) velocity, (e) panic factor $P$ and (f) averaged components of velocity and panic factor. The inset of (f) shows the instantaneous velocity profiles.
}\label{fig:figure3}
\end{figure}

\textit{Dynamic states and phase transition}: 
Two dynamically stable states of the system are obtained from our simulations. They are described by Fig~\ref{fig:figure3} (see SM \cite{sm1} for videos). 
At high values of the coordination coefficient $\mu$, active agents organize into an ordered velocity field (see Fig.~\ref{fig:half3p6}). The local orientation of this ordered velocity field is determined by the geometry of the confinement. In our simulations it manifests as a rotating velocity field induced by the circular confinement, a preferred state observed in other active systems as well \cite{wioland2013prl,lushi2014, czirok1996formation}. In periodic domains the ordered velocity field corresponds to a flocking state \cite{sm1} consistent with previous research \cite{chate2008collective}. From this state, as $\mu$ is reduced, collective motion disappears, the order in the velocity field is lost and we obtain a disordered state
(see Fig.\ref{fig:half3p4}). 

The two distinct states also  show different velocity profiles. In Fig.~\ref{fig:speed3p6}, the velocity field is dominated by its azimuthal component which increases from zero at the center to a maximum and reduces to zero again near the walls. On the other hand in Fig.~\ref{fig:speed3p4}, which corresponds to a disordered state, both the radial and azimuthal components of velocity fluctuate around zero mean and are small when compared to those observed under collective motion. This state where agents' motion is dictated by individual choice is referred to as the state of crush. In this state, the thrust force by each agent is balanced by the drag force, $m\beta \sim \mu d |\mathbf{v}_{i}|$
resulting in individual agent speed $|\mathbf{v}_{i}| \sim \frac{m \beta}{\mu d}$.

A brief discussion of the order-disorder transition is presented next.
Unlike previous variants of the Vicsek model \cite{vicsek1995prl,marchetti2013rmp1} 
where disorder is brought in via a random force, physical or virtual collisions (generated by elastic repulsive forces) acting in consonance with agent inertia is the source of disorder in our model. Activity further amplifies this noise in the system. An increase in coordination between agents reduces this deterministic noise \cite{sm1} and beyond a threshold value, we obtain ordered crowd motion. On the other hand, when noise dominates, the system goes into a state of crush.

In order to describe such a transition from an individual choice to a crowd dominated choice Helbing et al. \cite{vicsek2000nature} have postulated a \textit{panic factor}, $P$. Casting their \cite{vicsek2000nature} equations in our form, we identify the panic factor as 
\begin{align}
P =\frac{m \beta}{m \beta+\mu d \lvert \mathbf{v_c} \rvert}
\label{eqn:panic}
\end{align}
Thus panic factor is a measure of the individualistic behavior ratioed to the crowd behavior. In a panicked state, as observed in a crush, selfish active agents exhibit individualistic behavior and self propelling force will exceed the coordination force since $\mathbf{v_c} \approx 0$. Then $m\beta >> \mu d \lvert \mathbf{v_c} \rvert$ and $P \approx 1$. On the other hand, when the crowd is in an ordered state, the self-propelling force of individual agents would be just sufficient to match the forces exerted by the neighboring crowd: $m\beta \approx \mu d \lvert \mathbf{v_c}\rvert$ and $P \approx \frac{1}{2}$. Since $\mathbf{v_c}$ is a local quantity, Eq.~\eqref{eqn:panic} is  a generalization of the constant panic factor postulated by \cite{vicsek2000nature}. Thus, our model allows the crowd to have spatially varying choices. The panic factor corresponding to the ordered and disordered states are shown in Fig.~\ref{fig:panic3p6} and \ref{fig:panic3p4} respectively. As expected, $P\approx 1$ in the disordered state as opposed to $P \approx \frac{1}{2}$ in spatial regions of high order. The azimuthally averaged panic factor corresponding to the two states are also shown in Fig.~\ref{fig:speed3p6} and \ref{fig:speed3p4}. Notably, $\frac{1}{2} \le P \le 1$ in the Fig.\ref{fig:speed3p6}. 

In order to quantitatively describe these two states, we define the magnitude of the averaged velocity of the crowd, $\lvert\langle \mathbf{v} \rangle \rvert =  \frac{1}{T}{}\int_{0}^{T} \frac{1}{N}\sum_{i} \mathbf{v}_{i} dt$ as the order parameter.
%
%
Fig.~\ref{fig:figure2} shows the variation of $\lvert\langle \mathbf{v} \rangle \rvert$ as a function of $\mu$ for a fixed active forcing ($\beta =$ constant) and crowd density $\rho$. At large values of $\mu$, where collective motion dominates, the average velocity $\lvert\langle \mathbf{v} \rangle \rvert$, of the crowd is weakly dependent on $\mu$.
The collective motion persists even when $\mu$ is lowered but only upto a critical value, $\mu_1$. At $\mu_1$ there is a sharp decrease in the order parameter indicating a phase transition. The collective motion disappears and the agents exhibit diffusive motion. Further reduction in $\mu$ maintains $\lvert\langle \mathbf{v} \rangle \rvert \approx 0$. Reversing the experiment by increasing the value of $\mu$ gives a different $\mu_2$ for the reverse transition, namely order from a crush. 

 

\textit{Safety - Recovery from a crush via an induced transition}: So far, we analyzed the order - disorder transition of a crowd using $\mu$ as the control parameter. However, as Fig.~\ref{fig:figure2} indicates, the critical values for the transitions are different, $\mu_1 \ne \mu_2$. In other words, the system exhibits hysteresis. An interesting outcome of this observation is that two distinct stable dynamical states exist for $\mu_1 \le \mu \le \mu_2$. This presents an opportunity to investigate recovering order from a crush.

We start by considering a system exhibiting a state of crush for a particular $\mu$ in the hysteresis region, for example, the point marked as `$\times$' in Fig.~\ref{fig:figure2}. As can be seen, there also exists a state of order at this same level of coordination between the agents. We now empower game changers in the system and study whether the system can be driven to an ordered state. Game changers can be people either  employed at these locations a priori or they can be promoted agents. Game changers differ from the rest of the crowd only in their ability to control their velocity for a short duration of time. The distribution of game changers can be selected in different ways; we discuss two of them here: game changers located in a stripe at various radial locations (Fig.~\ref{fig:figure2}(a)) and uniformly dispersed throughout the domain (Fig.~\ref{fig:figure2}(b)).

In short, an impulse $\mathcal{I}=N_g m \gamma \tau$ is imparted to a state of crush through $N_g (=10\%N)$ game changers by assigning them a tangential motive force $m\gamma$ for a certain interval of time $\tau$. For the rest of the crowd, $\gamma=0$. The force is then removed and the system is allowed to evolve. The response of the system to varying impulse $\mathcal{I}$ corresponding to the point marked `$\times$' in Fig.~\ref{fig:figure2} is shown as an inset (c) in the same figure. It may be seen that $\lvert\langle \mathbf{v} \rangle \rvert$ steadily increases for $t > \tau$ when ${\mathcal{I}}/{\mathcal{P}_{o}}\geq 0.67$, indicating spontaneous restoration of order in the system after the forcing has been turned off. Here $\mathcal{P}_o=Nm{\lvert\langle \mathbf{v} \rangle \rvert}_o$ is the total momentum in the ordered state and $\lvert\langle \mathbf{v} \rangle \rvert_o$ is the average speed in the ordered state. 
On the other hand, when ${\mathcal{I}}/{\mathcal{P}_{o}}< 0.67$, $\lvert\langle \mathbf{v} \rangle \rvert$ reduces towards zero when the forcing is removed (for $t>\tau$). This implies that the system fails to recover from the crush despite the impulsive forcing.



Of course, successful recovery of the system from a state of crush depends upon (i) the strength of the impulse ($\mathcal{I}$) provided by game changers and (ii) the distribution of game changers - location, geometry and dimensions of the mode (for e.g.\ stripe vs dispersed). For $\mu_1 \le \mu \le \mu_2$, we have systematically identified the minimum value of $\mathcal{I}$ for which transition to order is achieved, for any possibility in $\frac{N_g}{N}$ and distribution of game changers. Thus we obtain the separatrices, in Fig.~\ref{fig:figure2} shown as dotted lines that the game changers have to move the system to, before spontaneous transition to the ordered state sets in.
When the system is impulsively forced to an order parameter value above (below) this curve, game changers succeed (fail) in recovering order. As $\mu$ increases, $\lvert\langle \mathbf{v} \rangle \rvert$ (and thus the corresponding impulse) required for the system to transit to an ordered state decreases. We also note that the separatrix that lies closest to the disordered state is for the case where the game changers are distributed on a ring located at $\approx 70\%$ radius of the domain. Correspondingly,  $\lvert\langle \mathbf{v} \rangle \rvert$ (and thus the corresponding impulse) required is always a minimum for this particular spatial distribution of game changers.

In other words, locating game changers at $70\%$ of the domain radius is the optimal way to recover from a crush in a circular domain without changing the level of coordination between the agents. This is explained by studying the radial velocity distribution as well as panic factor in the ordered state. At approximately $0.7R$, the azimuthal velocity ($|V_\theta|$) of the agents\, and therefore $\mathbf{v_c}$, are both at their maxima, as seen in Fig.~\ref{fig:speed3p6}. Hence the power input to the system ($\mathcal{I}|\mathbf{v_c}|$) through the organized motion of the game changers is maximum. Therefore, the impulse required to transform the system is minimum when they are placed at the locations of maximum crowd speed. Maximum $|\mathbf{v_c}|$ at this location also implies that the panic factor (and thus individualism) is at a minimum. 

This is an important result with regard to crowd safety. In order to recover from a crush without changing the level of coordination between the agents while imparting the least impulse, game changers would have to be placed at the locations of maximum crowd speed which correspond to locations of lowest panic in the domain. In case of an accident like crush, these game changers can impart power to the crowd in identified directions, thus helping align agent motion throughout the domain.

We have also studied the transition from an ordered state to a disordered state. Here the role of game changers is observed to be similar to the role of a small fraction of rogue elements in flocking models studied earlier \cite{Ariel2015, yllanes2017many}. We find that breaking order via rogue elements is easier than its restoration.

\begin{figure}
\centering
\subfigure[]{{\includegraphics[scale=0.17]{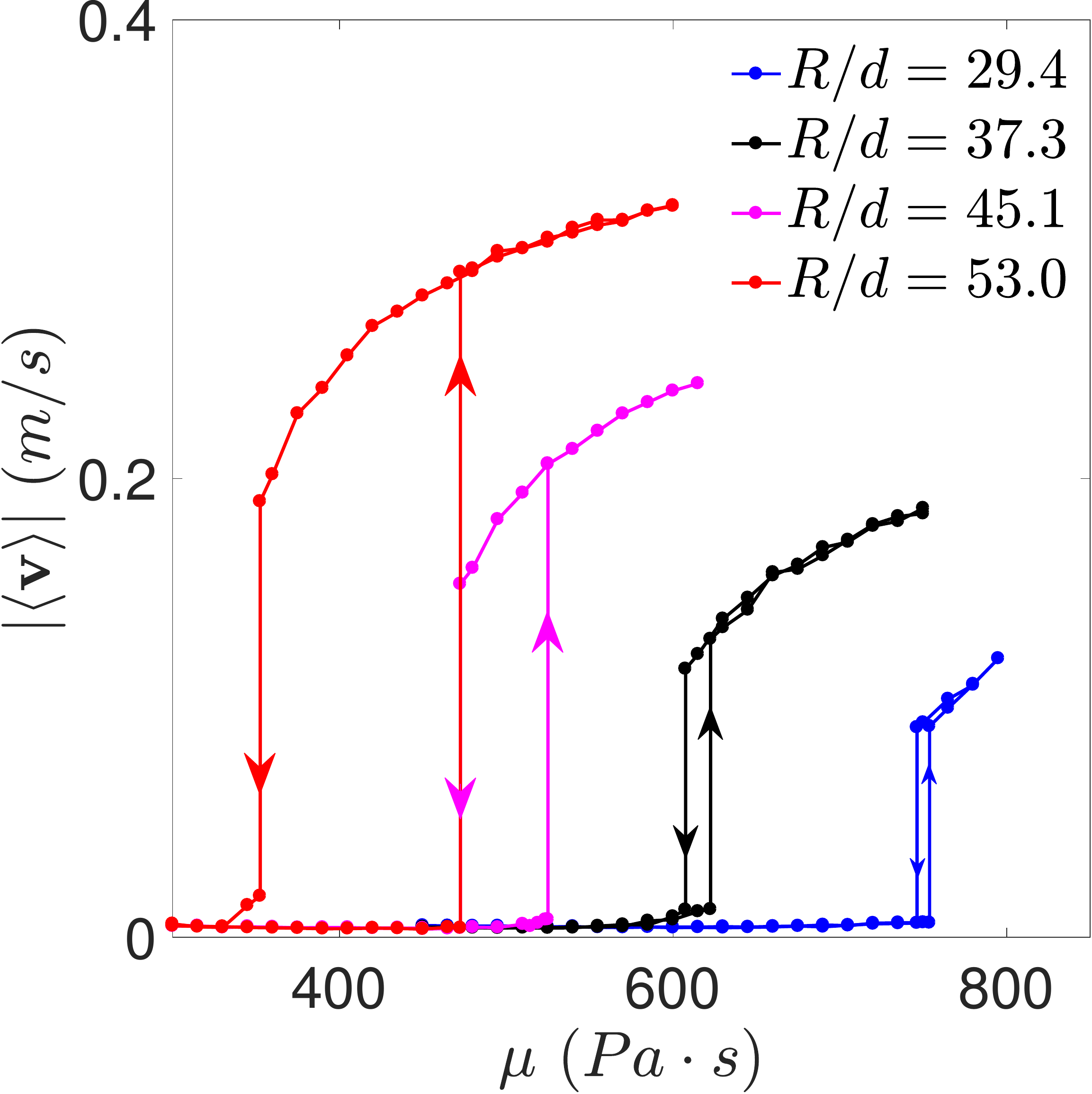}\label{fig:orderp}}}
\subfigure[]{\includegraphics[scale=0.16]{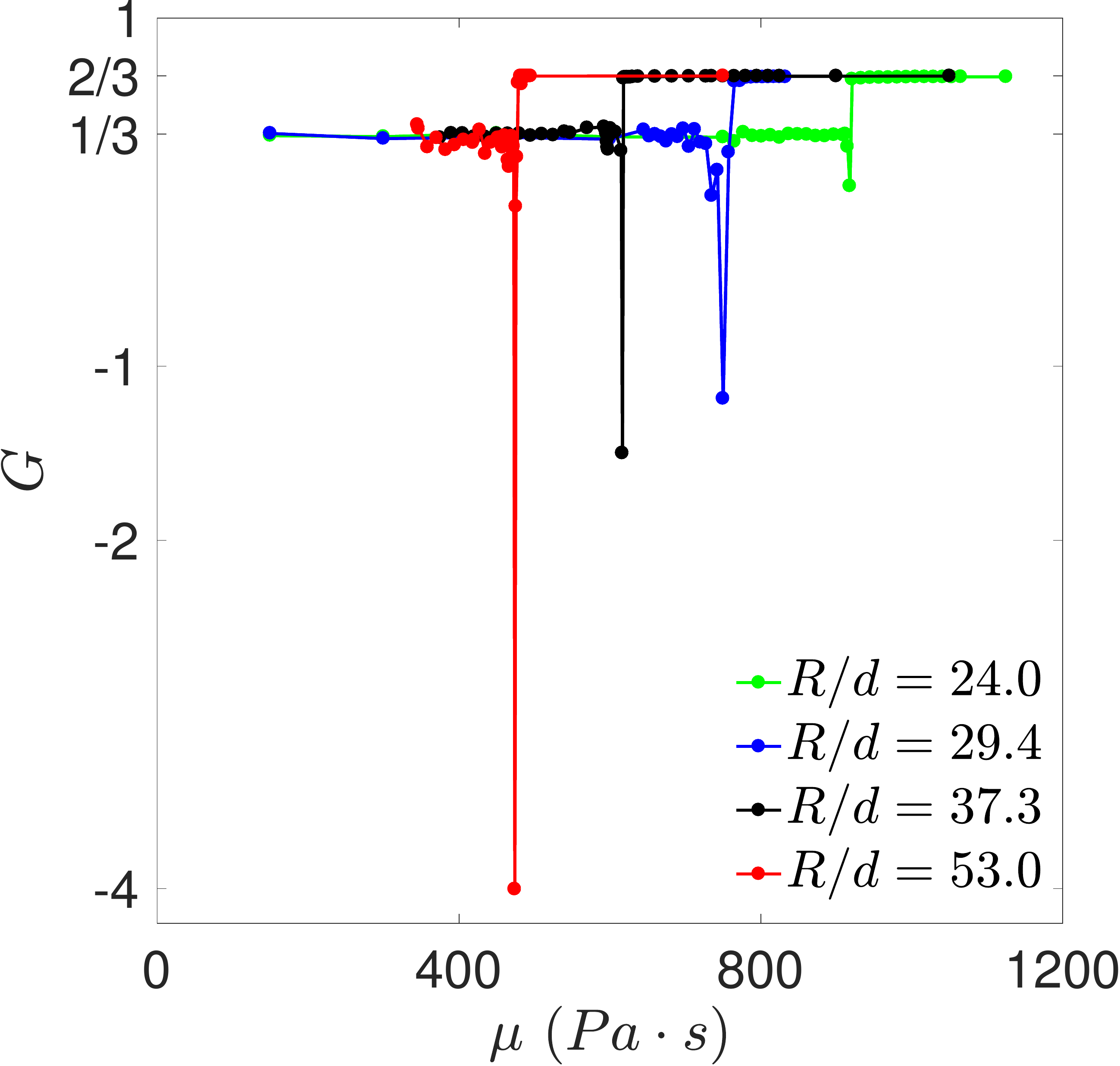}\label{fig:binder}}
\caption{(a) During the order-disorder transition in a crowd, the order parameter $\lvert\langle \mathbf{v} \rangle \rvert$ shows a discontinuous change at $\mu = \mu_1, \mu_2$. Different sets correspond to different system sizes. (b) Change in the value of Binder cumulant from $1/3$ to $2/3$ through a dip at $\mu = \mu_2$ during the transition.}

\end{figure}

\textit{First order phase transition and hysteresis:} We now discuss the reasons that lead to the basic premise that the recovery from a crush via game changers is possible and we show that this possibility exists due to the underlying nature of the phase transition.
The sharp change in the order parameter in Fig.~\ref{fig:orderp} at $\mu=\mu_1, \mu_2$ indicates that the phase transition from an ordered state to a crush is first order. We confirm this by calculating the Binder cumulant of the order parameter through the transition while moving to a state of order from a state of crush. We define the Binder cumulant $G$ following Chat\'e et al. \cite{chate2008collective} and plot it as a function of the control parameter $\mu$. 
%
%
Fig.~\ref{fig:binder} shows this plot for various domain sizes, $R$, for a fixed $\rho$. $G$ varies from  $2/3$ to $1/3$ from an ordered state to a disordered state as is known for two dimensional systems.
During the transition, Binder cumulant undergoes a sharp dip at the transition point proving that crowd crush is indeed a first order phase transition from a uniformly moving crowd. Moreover the size and the sharpness of the dip also increases with an increase in the system size, another feature of a first order phase transition \cite{chate2008collective}. Thus, our simulations underscore the discontinuous nature of order-disorder transition seen in Vicsek-like models of active systems \cite{chate2008collective}, additionally under confinement.

Hysteresis is a hallmark of first order phase transitions. Therefore, near the transition point,  both ordered and disordered states are solutions at the same $\mu$. This existence of multiple solutions is the reason that opens a pathway to reverse a crush state through mobilization of game-changers. It appears that the role played by game changers in restoring order is analogous to nucleation in equilibrium phase transitions \cite{binder1987theory,hohenberg1995metastability,csernai1992nucleation}. Our analysis also shows that it would not be possible to recover the state through game changers for $\mu < \mu_1$. On the other hand, the width of the hysteresis loop increases with an increase in domain size (see Fig.~\ref{fig:orderp}), providing better scope for recovery in case of an accident in large gatherings.



\begin{acknowledgments}
The authors would like to gratefully acknowledge discussions with Srikanth Vedantam, Silke Henkes and Hugues Chat\'e.
\end{acknowledgments} 

\bibliography{prl.bib}




\end{document}